\documentclass[twocolumn,showpacs,preprintnumbers,amsmath,amssymb,groupedaddress]{revtex4}
\usepackage{graphicx}
\usepackage{dcolumn}
\usepackage{bm}

\begin{document}


\title{Hidden symmetry and Collective behavior}

\author{Bin Ao}
\author{Zhigang Zhu}
\author{Liang Huang}
\author{Lei Yang\cite{ca}}

\affiliation{Institute of Modern Physics, Chinese Academy of
Science, Lanzhou 730000, China}
\date{\today}

\begin{abstract}
We study the relationship between the partially synchronous state
and the coupling structure in general dynamical systems. Our results
show that, on the contrary to the widely accepted concept,
topological symmetry in a coupling structure is the sufficient
condition but not the necessary condition. Furthermore, we find the
necessary and sufficient condition for the existence of the partial
synchronization and develop a method to obtain all of the existing
partially synchronous solutions for all nonspecific dynamics from a
very large number of possible candidates.
\end{abstract}

\maketitle

Synchronization has attracted extensive attention in the physical,
biological, ecological, and other systems \cite{application}.  The
theory of synchronization focuses on the dynamical behavior of
many-body coupled systems. Of the types of synchronization, global
synchronization (GS) is a well researched topic \cite{gs}. Partial
synchronization (PaS), which is the synchronization only emerges in
part of a system, is a more general synchronous phenomenon. And PaS
has been observed in the systems whose parameters are outside the GS
regime. A ``simple" case, the PaS in a globally coupled system, is
studied in detail in Refs. \cite{kaneko, amritkar_hu}. And random
and other complex coupling structures are studied in Refs.
\cite{symm_manrubia, amritkar_hu, other_pas}. However, the
underlying mechanism of PaS is yet far from clear. A fundamental
difficulty lies in finding the PaS solutions for a given structure.
And the problem remains open.

\begin{figure}[tbp]
\includegraphics[width=3in,height=1.2in]{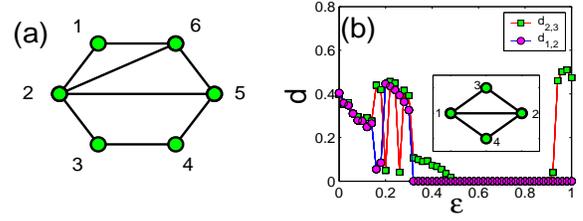}
\caption{(a) A topological structure of the coupled system without
any partial synchronous solution. (b) The average distance
$d_{1,2}$ and $d_{2,3}$ versus the coupling strength $\varepsilon
$ with the structure shown in the inset graph.}
\label{nonexistence_n4}
\end{figure}

Considering the dynamics \cite{sys}
\begin{eqnarray}
\textbf{X}_{n+1}=\textbf{F}(\textbf{X}_n)+\varepsilon
(C\otimes\Gamma )\textbf{F}(\textbf{X}_n), \label{map}
\end{eqnarray}
where $\mathbf{X}=(\mathbf{x}^{1},\mathbf{x}^{2},\cdots
,\mathbf{x}^{N})'$ represents the state of a system consisted of $N$
subsystems $\{\mathbf{x}^{i}\}_{i=1}^N$ and "$()'$" means the
transpose of a matrix. The independent state, $\mathbf{x}^{1}\neq
\mathbf{x}^{2}\neq \cdots \neq \mathbf{x}^{N}$, and the GS state,
$\mathbf{x}^{1}=\mathbf{x}^{2}=\cdots =\mathbf{x}^{N}$, universally
exist for common couplings \cite{couple}. The PaS solution, however,
may not exist in a given coupled system. As an example, for the
structure shown in Fig. \ref{nonexistence_n4}(a), all 203 candidate
PaS solutions (e.g., $\mathbf{x}^1=\mathbf{x}^2$ and
$\mathbf{x}^3\neq\mathbf{x}^4\neq\mathbf{x}^5\neq\mathbf{x}^6$; this
is a kind of exhaust algorithm) do not satisfy Eq. (\ref{map}).

In recent years, it has been developed to a common belief that
there exist close relation between the topological symmetry of a
coupling structure and the PaS state. For example, an asymmetric
PaS pattern that does not follow a symmetrical structure has never
been observed \cite{symm_manrubia}, the theory of symmetric group
can be used to describe the periodic PaS state in several regular
structures with the same symmetry \cite{symm_yu}, and all PaS
states corresponding to each topological symmetry in a ring have
been observed \cite{symm_zhang}. Based on these examples, one
could suppose that symmetry is the necessary and sufficient
condition for the existence of the PaS state.

When the number of subsystems is small, the above statement may seem
to hold true. The coupling structure in the inset of Fig.
\ref{nonexistence_n4}(b) given by the adjacency matrix
$\begin{array}{cc} A_{4}=\left(
\begin{array}{cccc}
0 & 1 & 1 & 1 \\
1 & 0 & 1 & 1 \\
1 & 1 & 0 & 0 \\
1 & 1 & 0 & 0%
\end{array}%
\right). &
\end{array}$
Here, nodes 1 and 2 are symmetric, and $A_4$ is invariant under the
permutations $1\rightarrow2$ and $2\rightarrow1$. The curves
$d_{1,2}(\varepsilon)$ and $d_{2,3}(\varepsilon)$ \cite{aver_dis}
are shown in Fig. \ref{nonexistence_n4} (b), where $\varepsilon$ is
the coupling strength. The synchronous solution
$\mathbf{x}^1=\mathbf{x}^2\neq\mathbf{x}^3\neq\mathbf{x}^4$ is
observed in the region $\varepsilon \in [0.3,0.45]\cup[0.9,1]$.
Thus, the PaS state will be achieved with the corresponding symmetry
in $A_4$ among the symmetrical nodes.

\begin{figure}[tbp]
\includegraphics[width=3in, height=1in]{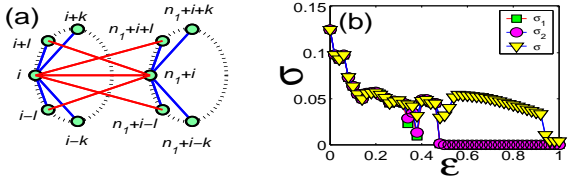}
\caption{(a) The scheme of a topological structure with the
"rotational" symmetry. (b) The variation of the two clusters
$\sigma_1$, $\sigma_2$ and the whole system $\sigma$ as functions of
$\varepsilon$ with $n_{1}=n_{2}=100$, $k=40$, and $l=10$.}
\label{symm_example1}
\end{figure}

A more complex case is shown in Fig. \ref{symm_example1} (a) with
the same dynamics as above. There are two clusters and their nodes
are denoted $ 1,2,\cdots ,n_{1}$ and $n_{1}+1,n_{1}+2,\cdots,
n_1+n_2$. Node $i$ $(1\leq i\leq n_1)$ is coupled to nodes $i-k,
i-k+1, \cdots, i+k$, $ n_1+i-l, n_1+i-l+1, \cdots, n_1+i+l$; and
node $n_1+i$ $(1\leq i\leq n_1)$ is coupled to the nodes $n_1+i-k$,
$n_1+i-k+1, \cdots$, $n_1+i+k$, $i-l, i-l+1, \cdots, i+l$.
Obviously, there is "rotational" symmetry in every cluster; in other
words, the adjacency matrix is invariant under a "rotation"
permutation in each cluster; that is, $1\rightarrow 2$,
$2\rightarrow 3$, $\cdots$, $n_{1}-1\rightarrow n_{1}$,
$n_{1}\rightarrow 1$, and $n_1+1\rightarrow n_1+2$,
$n_1+2\rightarrow n_1+3$, $\cdots$, $2n_{1}-1\rightarrow 2n_{1}$,
$2n_{1}\rightarrow n_1+1$. We define a $M\times M$ matrix $R_{M}$,
where $(R_{M})_{M,1}=1$, $(R_{M})_{i,i+1}=1$ $(i=1,2,\cdots ,M-1)$,
and the other elements are 0. Thus, the permutation matrix
\cite{perm_matr} of this transformation will be
$T_{d}=R_{n_{1}}\oplus R_{n_{2}}$, where $\oplus$ represents the
direct sum of two matrices. Fig. \ref{symm_example1} (b) shows the
time-averaged variation in all subsystems $\sigma(\varepsilon)$, and
that in the two clusters, $\sigma_1(\varepsilon)$ and
$\sigma_2(\varepsilon)$ \cite{aver_dis} for $n_{1}=$ $n_{2}=100$,
$k=40$, and $l=10$. For $\varepsilon\in[0.45,1]$,
$\sigma_1=\sigma_2=0$; that is,
$\mathbf{x}_1=\cdots=\mathbf{x}_{n_1}$ and
$\mathbf{x}_{n_1+1}=\cdots=\mathbf{x}_{N}$. The PaS solution of the
"rotational" symmetrical nodes is observed. The above examples show
that the presence of symmetry in a structure may be necessary and
sufficient for the existence of PaS solutions.

In this Letter, we investigate in detail the relationship between
the PaS solution and the coupling structure. The PaS solution is
defined as follows: For a dynamical system with phase space
$\mathbb{R}^{Nm}$, a $K$-cluster synchronous solution is a
$Km$-dimensional subspace $V$ of $\mathbb{R}^{Nm}$. It can be
represented by
\begin{equation}
\begin{array}{c}
\mathbf{x}^{i_{1}^{1}}=\mathbf{x}^{i_{1}^{2}}=\cdots =\mathbf{x}%
^{i_{1}^{n_{1}}}, \\
\mathbf{x}^{i_{2}^{1}}=\mathbf{x}^{i_{2}^{2}}=\cdots =\mathbf{x}%
^{i_{2}^{n_{2}}}, \\
\cdots \cdots \\
\mathbf{x}^{i_{K}^{1}}=\mathbf{x}^{i_{K}^{2}}=\cdots =\mathbf{x}%
^{i_{K}^{n_{K}}},%
\end{array}
\label{syn_sta_1}
\end{equation}
where $\mathbf{x}^{i_{a}^{b}}$ denotes the $b$th subsystem in the
$a$th cluster; and $\{n_{i}\}_{i=1}^{K}$ are the sizes of each
cluster, where $\sum_{i=1}^{K}n_{i}=N$ ($N>K$ $>1$). The GS and
independent solutions are the particular cases where $K=1$ and
$K=N$, respectively. The relationship between the PaS solution and
the coupling structure could be described by the following two
questions:

Question A: If one finds symmetry in a coupling structure, can a
corresponding PaS solution be obtained?

Eq. (\ref{syn_sta_1}) can also be described by its matrix form:
\begin{equation}
(T\otimes I_m)\mathbf{X}=\mathbf{X},\forall \mathbf{X}\in V,
\label{syn_sta}
\end{equation}
where $T$ is a permutation matrix and $I_m$ is a $m$-dimension
identity matrix. That is, $\mathbf{X}\in V$ is invariant under the
permutation transformation $T\otimes I_m$, so $V$ is the invariant
subspace of $T\otimes I_m$ and the eigen-subspace of $T\otimes I_m$
corresponding to eigenvalue 1. That $V$ is the invariant subspace of
the dynamical system in Eq. (\ref{map}) requires
\begin{equation}
(C\otimes\Gamma)\mathbf{X}\in V,\ \ \ \forall \mathbf{X}\in V.
\label{solution}
\end{equation}
Next, if there is symmetry $T$ in structure $C$, then $C$ will be
invariant under the permutation transformation $T$. The mathematical
representation is
\begin{equation}
T^{-1}CT=C,  \label{symm_net}
\end{equation}
or the matrices $C$ and $T$ are commutative in multiplication.

Question A can be represented by the following mathematical
statement:
$$
\begin{array}{cc}
If \ Eq.\ (\ref{syn_sta})\ and \  Eq.\ (\ref{symm_net})\  hold, &
then \ Eq.\ (\ref{solution})\  holds.
\end{array}
$$
Multiplying the two sides of Eq. (\ref{symm_net}) by $\mathbf{X}\in
V$ and arbitrary $\Gamma$, we have
\begin{equation}
\begin{array}{c}
[(TC)\otimes\Gamma]\mathbf{X}=[(CT)\otimes\Gamma]\mathbf{X},\forall
\mathbf{X}\in V.
\end{array}
\label{proof_suf_1}
\end{equation}
Combining Eq. (\ref{syn_sta}) with Eq. (\ref{proof_suf_1}) gives
\begin{equation}
(T\otimes I_m)(C\otimes \Gamma)\mathbf{X}=(C\otimes
\Gamma)\mathbf{X},\forall \mathbf{X}\in V. \label{proof_suf_2}
\end{equation}
Thus, $(C\otimes \Gamma)\mathbf{X}$ is also the eigenvector of
$T\otimes I_m$ with eigenvalue 1. Then, Eq. (\ref{solution}) will be
satisfied for all $\mathbf{X}\in V$. We conclude that for a
symmetrical structure, the dynamical system has a corresponding PaS
solution.

Question B: If one finds a PaS solution, can the corresponding
symmetry in the coupling structure be observed?

\begin{figure}
\includegraphics[width=3in,height=1in]{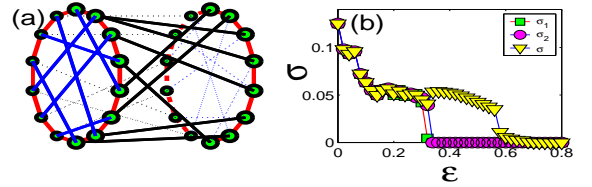}
\caption{(a) The scheme of an equal degree random structure. (b) The
variance of the two clusters $\protect\sigma _{1}$, $\protect \sigma
_{2}$ and the whole network $\protect\sigma $ as functions of the
coupling strength $\protect\varepsilon$ in the parameter
$n_{1}=n_{2}=100$, $p=1$, and $p_{r}=0.5$. } \label{edr}
\end{figure}

Here, an interesting example is shown in Fig. \ref{edr} (a). Let us
consider two clusters; each has $n$ subsystems and every subsystem
is randomly connected to $[pn]+1$ subsystems in the same cluster
($p$ is a probability and $[pn]$ means the integer part of $pn$) and
$[p_{r}n]+1$ subsystems in the other cluster ($p_{r}$ is also a
probability). All the subsystems are of equal degree and the
connections between two clusters are also of equal degree. Fig.
\ref{edr} (b) shows the variance of the two clusters ($\sigma _{1}$
and $\sigma _{2}$) and of the whole system ($\sigma $) as functions
of the coupling strength $\varepsilon $ for $n=100$, $p=1$, and
$p_{r}=0.5$. As $\varepsilon \in \lbrack 0.34,0.76 \rbrack$, $\sigma
_{1}=$$ \sigma _{2}=0$, and $\sigma >0$, the PaS solution is
observed. Due to the random connections between the subsystems,
there is no symmetry in the structure.

Question B can be represented by the following mathematical
statement:
$$
\begin{array}{cc}
If \ Eq.\ (\ref{syn_sta})\ and \ Eq.\ (\ref{solution})\  hold, &
then \ Eq.\ (\ref{symm_net})\ holds.
\end{array}
$$
The following relations could be derived:
\begin{eqnarray}
&(C\otimes\Gamma)\mathbf{X}\in V,(T\otimes I_m)\mathbf{X}=\mathbf{X},\forall \mathbf{X}\in V, \notag \\
\Longleftrightarrow &(T\otimes I_m)(C\otimes\Gamma)\mathbf{X}=(C\otimes\Gamma)\mathbf{X},  \notag \\
\Longleftrightarrow &(T\otimes I_m)(C\otimes\Gamma)\mathbf{X}=(C\otimes\Gamma)(T\otimes I_m)\mathbf{X},  \notag \\
\Longleftrightarrow &[(TC)\otimes(I_m\Gamma)]\mathbf{X}=[(CT)\otimes(\Gamma I_m)]\mathbf{X},  \notag \\
\Longleftrightarrow &[(TC-CT)\otimes\Gamma]\mathbf{X}=0.
\label{solution_exist}
\end{eqnarray}
Obviously, Eq. (\ref{solution_exist}) is not equivalent to Eq.
(\ref{symm_net}). We can conclude that it is possible for a PaS
solution to exist in a dynamical system without any symmetry; Fig.
\ref{edr} presents an example. Another important point is that, in
fact, the necessary and sufficient condition can be drawn from Eq.
(\ref{solution_exist}) itself.

For a simple representation but not loss its generality, we only
discuss the case $m=1$. The component form of Eq.
(\ref{solution_exist}) is
\begin{equation}
\sum_{j=1}^{N}(C_{kj}-C_{ij})\mathbf{x}^{j}=0, \forall
\mathbf{X}=(\mathbf{x}^{1},\mathbf{x}^{2},\cdots,\mathbf{x}^{N})\in
V. \label{solution_exist_comp}
\end{equation}
for $T_{ik}=1$. By relabeling the subsystems to group subsystems of
the same cluster together, we rewrite Eq. (\ref{syn_sta_1}) as
below: For cluster $s$,
\begin{equation}
\begin{array}{c}
\mathbf{x}^{N_s+1}=\mathbf{x}^{N_s+2}=\cdots =\mathbf{x}^{N_s+{n_{s}}%
}\equiv \mathbf{y}^{s}, \ (s=1,2,\cdots,K),
\end{array}
\label{syn_sta_2}
\end{equation}
where $N_s=0$ when $s=1$, and $N_{s}=\sum_{i=1}^{s-1}n_{i}$ when
$s=2,\cdots,K$ \cite{t4td}. Then Eq. (\ref{solution_exist_comp}) can
be grouped into $K$ terms as
\begin{equation}
\sum_{s=1}^K[\sum_{j=N_{s}+1}^{N_{s}+n_{s}}(C_{ij}-C_{kj})]\mathbf{y}^{s}=0,\
(i=1,2,\cdots,N).\label{equ_deg_sum}
\end{equation}
This is the $i$th row of Eq. (\ref{solution_exist}). For all
nonspecific dynamics, we always have
\begin{equation}
\sum_{j=1}^{n_{s}}(C_{i,N_{s}+j}-C_{k,N_{s}+j})=0,(
s=1,2,\cdots,K)\label{equ_deg}
\end{equation}
for $T_{ik}=1$($i=1,2,\cdots,N$).

For a subsystem $i$ which belongs to cluster $s$, we define the
degree of subsystem $i$, which is contributed by the $s'$th cluster
($s'\neq s$), as the \textbf{external degree} of $i$ from $s'$; And
the degree of subsystem $i$, which is contributed by the $s$th
cluster as the \textbf{internal degree} of $i$. Since $C_{ij}$ may
be a fraction or zero, the external and internal degrees of a
subsystem can also be fraction or zero.

For subsystem $i$ in cluster $s$,
$\sum_{j=1}^{n_{s'}}C_{i,N_{s'}+j}$ is the external degrees of $i$
from $s'$. And because of $T_{ik} = 1$, both subsystems $i$ and $k$
are in cluster $s$. Then from Eq. (\ref{equ_deg}), the external
degrees of $i$ and $k$ from $s'$ should be the same. And,
$\sum_{j=1}^{n_{s}}C_{i,N_{s}+j}-C_{ii}$ is the internal degrees of
$i$. Because of $C_{kk}=C_{ii}=-1$, the rest part of Eq.
(\ref{equ_deg}) shows that the internal degrees of $i$ and $k$
should be the same.

Considering the above two different situations in Eq.
(\ref{equ_deg}), we can have the following necessary and sufficient
conditions of the existence of PaS solutions: \textsl{\textbf{S1}}:
\textbf{the external degrees of every subsystems in a cluster from
another cluster should be the same.} \textsl{\textbf{S2}}:
\textbf{the internal degrees of the subsystems in a cluster should
also be the same}.

The two conditions form a complete representation of Eq.
(\ref{solution_exist}). Now we have a clear physical picture of the
existence of a PaS solution. And from the two conditions, we can
derive all the PaS solutions in a given structure using the
following strategy: The key is to divide the structure into several
substructures, each comprising subsystems that can synchronize with
each other but not with subsystems in other substructures. Then the
PaS solutions of the whole system will be combinations of all the
possible solutions of every substructure.

To satisfy \textbf{S1} and \textbf{S2}, the PaS cannot be achieved
between directly connected subsystems whose degrees are coprime
\cite{coprime}. So an $N\times N$ matrix $S$ can be defined such
that $S_{im}=1$ if the degrees of the subsystems $i$ and $m$ are not
coprime or $C_{im}=0$, and $S_{im}=0$ for other cases. Thus, the
system can be divided into several groups of subsystems, and if the
subsystems $i$ and $m$ are in different groups, then $S_{im}=0$. We
can then suppose that $\textbf{x}^i=\textbf{x}^m$ when $S_{im}=1$.
Thus, an equation like Eq.(\ref{solution_exist_comp}) can be
represented as $\sum_{j_+\in
J_+}C^+_{j_+}\textbf{x}^{j_+}+\sum_{j_-\in
J_-}C^-_{j_-}\textbf{x}^{j_-}=0$, where $j_+\in
J_+=\{j|C^+_{j}\equiv C_{mj}-C_{ij}>0\}$ and $j_-\in
J_-=\{j|C^-_{j}\equiv C_{mj}-C_{ij}<0\}$. Positive (negative)
coefficients $C^+_{j_+}$ ($C^-_{j_-}$) indicate that the internal or
external degrees of subsystem $m$ from subsystem $j_+$ ($j_-$) is
greater (less) than that of the subsystem $i$ from $j_+$ ($j_-$).
Therefore, according to the requirement of \textbf{S1} and
\textbf{S2}, subsystem $j_+$ should belong to a cluster that
includes one or more subsystems in $J_-$ so that the positive degree
difference $C^+_{j_+}$ can be counteracted. But subsystem $j_+$ may
not be synchronous with some members of $J_-$ (some elements
$S_{j_+j_-},\ \forall j_-\in J_-$ may be 0); if
$C^+_{j_+}+\sum_{j_-\in J_-}S_{j_+j_-}C^-_{j_-}>0,\exists j_+\in
J_+$, then Eq.(\ref{solution_exist_comp}) cannot be satisfied. This
means $\textbf{x}^i=\textbf{x}^m$ is impossible and $S_{im}$ should
be reset to 0. For the reasons specified above, if
$C^-_{j_-}+\sum_{j_+\in J_+}S_{j_-j_+}C^+_{j_+}<0,\exists j_-\in
J_-$, $S_{im}$ should be reset to 0. To let $i,m=1,\cdots,N$ and
perform the operations discussed above, we obtain a new $S$. Thus,
the system can be divided into some subgroups by considering the new
$S$. This procedure can be repeated until the new $S$ equals the old
$S$, and the size of the subgroups of the subsystems will decrease
with every repetition. After repeated subdivision, we obtain a
number of final groups, which can be grouped into clusters using the
exhaust algorithm. If S1 can be satisfied for every pair of clusters
and S2 can be satisfied in each individual cluster, combinations of
these clusters are PaS solutions of the whole system.

\begin{figure}
\includegraphics[width=3.4in,height=1.4in]{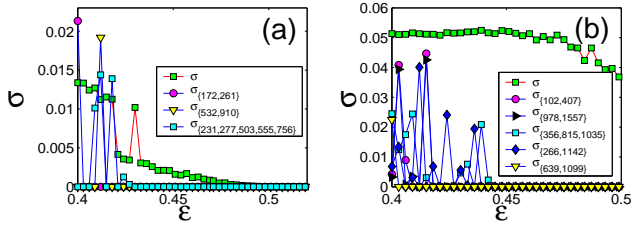}
\caption{The variance of the PaS clusters and the whole system
$\sigma$ as functions of the coupling strength $\varepsilon$ (a) A
random structure: $N=1000$, $p=0.995$. (b) A Barab\'{a}si-Albert
scale free structure: $N=1600$. } \label{pas_solution}
\end{figure}

The number of candidate PaS solutions for any structure rapidly
increases with N \cite{solution_numbers}. Thus, for a random
structure with $N = 1000$ and $p = 0.995$ \cite{matrix}, it would
be impossible to test all the candidates. But we can use the above
procedure to find all PaS solutions. Three equal-degree groups
that include at least two subsystems are found: $group\ 1,
\{172,261\}$; $group\ 2,\{532,910\}$; and $group\
3,\{231,277,503,555,756\}$. Thus, there are $2\times2\times52=208$
different PaS solutions. Fig. \ref{pas_solution}(a) shows the
variances of the whole system, $\protect\sigma $, and the three
clusters, $\protect\sigma _1$, $\protect \sigma _2$, and $\protect
\sigma _3$, as functions of the coupling strength
$\protect\varepsilon$. It is easy to find the PaS solutions
$x_{172}=x_{261}$, $x_{532}=x_{910}$, and
$x_{231}=x_{277}=x_{503}=x_{555}=x_{756}$. For the case of a
Barab\'{a}si-Albert scale free structure ($N=1600$) \cite{ba,
matrix}, Fig. \ref{pas_solution}(b) show the existence of PaS
clusters $\{102,407\}$, $\{978,1557\}$, $\{356,815,1035\}$,
$\{266,1142\}$, and $\{639,1099\}$ \cite{matrix}.

In conclusion, we studied the relationship between the coupling
structure and the PaS state in general dynamical systems. The
necessary and sufficient condition of the existence of a PaS state
was found from an exact proof. And the results are
counterintuitive; the existence of a PaS state does not require
symmetry, as assumed previously. And all of the candidate PaS
solutions exist in a globally coupled system. Furthermore, as the
exhaust algorithm cannot be used to obtain all of the existent PaS
solutions, we developed a method to find all these solutions for a
given structure. We focused mainly on the existence of PaS
solutions in this letter. But to determine the stability of these
solutions, the conditional Lyapunov exponent should be calculated
by regarding the PaS manifold as the condition. Note that the
proof also can also be applied to differential dynamical systems
and the conclusions are the same.

We are grateful for discussions with Ping Lin, Jinfang Zhang, Ying
Hu and Wei Kang. This work was supported in part by the 100 Person
Project of the Chinese Academy of Sciences.


\begin{thebibliography}{99}
\bibitem{sys} This is one of the widely used forms of the coupling nonlinear maps.
One can use a certain $m$-dimensional map $\mathbf{x}_{n+1}=
\mathbf{f}(\mathbf{x}_{n})$, $\mathbf{x}=(x_{n}^{1},x_{n}^{2},\cdots
,x_{n}^{m}) $ as a subsystem to construct a coupled system
$\mathbf{X}=(\mathbf{x}^{1},\mathbf{x}^{2},\cdots
,\mathbf{x}^{N})'$,
$\mathbf{F}=(\mathbf{f}^{1},\mathbf{f}^{2},\cdots
,\mathbf{f}^{N})'$. The common interest is linear coupling between
the subsytems; thus, the coupling term will be $\varepsilon
C\otimes\Gamma\mathbf{F}(\mathbf{X})$, where $\varepsilon $ the
coupling strength. $\Gamma :\mathbb{R}^{m}\rightarrow
\mathbb{R}^{m}$ characterizes the coupling scheme among the
subsystems and $C$ is the Laplacian matrix. $C_{ij}=A_{ij}/k_{i}$
for $j\neq i$ and $C_{ii}=-1$, where $k_{i}$ is the degree of
subsystem $i$ and $A_{ij}$ is an element of the adjacency matrix $A$
of the coupling structure. To gain further insights in numerical
simulation, we set $f(x)$ as be the logistic map $f(x)=4x(1-x)$, but
the proof does not depend on the dynamics.
\bibitem{aver_dis} One can define the average distance
between two oscillators $i$ and $j$ by $d_{ij}=\lim_{N\rightarrow
\infty }(N^{-1}\sum_{n=1}^{N}\Vert
\mathbf{x}_{n}^{i}-\mathbf{x}_{n}^{j}\Vert )$ and the time-averaged
variation among subsystems
$\sigma=\lim_{N'\to\infty}\frac{1}{N'}\sum_{n=1}^{N'}\sqrt{\sum_{i=1}^N(\mathbf{x}^i_n-\frac{1}{N}\sum_{i=1}^N\mathbf{x}^i_n)^2}$,
and state that they are in a synchronous state in the sense that
$d_{ij}<10^{-12}$.
\bibitem{t4td} The general form of $T$ will be
$T=\bigoplus_{i=1}^{K}R_{n_{i}}\ (i=1,2,\cdots ,K).$ Then the
general form of $\mathbf{X}$ will be $\mathbf{X}=(\mathbf{y}
^{1}\otimes1_{1,n_{1}},\mathbf{y}^{2}\otimes1_{1,n_{2}},\cdots
,\mathbf{y}^{K}\otimes1_{1,n_{K}})'$ where $1_{M,N}$ is an $M\times
N$ matrix in which every element is 1. Then the $T_{d}$ in the
second example is only a special case of this form, and the
permutation transformation done to $A_4$ can also be represented by
$T_{4}=R_{2}\oplus 1\oplus 1$.
\bibitem{solution_numbers} For example, there
are only 52 possible solutions (including the GS and the independent
solutions) in a structure with five subsystems, but the number will
be more than $1.38\times10^9$ when the system size is just 15.
\end{thebibliography}
\end{document}